\documentclass[pra,twocolumn,showpacs]{revtex4}
\usepackage{graphicx,amsfonts,amsmath,amssymb}
\usepackage{times}

\begin{document}

\title{Polarization spectroscopy in rubidium and cesium}
\author{M. L. Harris, C. S. Adams, S. L. Cornish, I. C. McLeod, E. Tarleton and I. G. Hughes}
\affiliation{Department of Physics, University of Durham, Rochester
Building, South Road, Durham DH1 3LE, United Kingdom}

\date{\today}

\begin{abstract}
We develop a theoretical treatment of polarization spectroscopy
and use it to make predictions about the general form of
polarization spectra in the alkali atoms. Using our model, we
generate theoretical spectra for the D2 transitions in $^{87}$Rb,
$^{85}$Rb, and $^{133}$Cs. Experiments demonstrate that the model
accurately reproduces spectra of transitions from the upper
hyperfine level of the ground state only. Among these, the closed
transition $F\rightarrow F'=F+1$ dominates, with a steep gradient
through line center ideally suited for use as a reference in laser
locking.
\end{abstract}


\maketitle

\section{Introduction}

Polarization spectroscopy is a sub-Doppler spectroscopic technique
in which birefringence is induced in a medium by a circularly
polarized pump beam, and interrogated with a counterpropagating weak
probe beam \cite{Wieman,Teets,Demtroder}. It is closely related to
saturation spectroscopy, and has proved a useful tool in areas as
diverse as combustion diagnostics \cite{Eckbreth}, plasma
characterization \cite{Danzmann}, and laser frequency stabilization
\cite{Pearman, Ratnapala, Streed}.

In this work we develop a simple theoretical model of polarization
spectroscopy based on a calculation of rate equations. Using this
model, we generate spectra for the hyperfine transitions of the D2
lines of rubidium and cesium, and show that for transitions from the
upper hyperfine level of the ground state ($n \:^2S_{1/2}\:
(F=\mathcal{I}+1/2)\rightarrow n \:^2P_{3/2} \:(F'=F+1, F,F-1$,
where $\mathcal{I}$ is the nuclear spin) our model gives good
agreement with the observed spectra. For transitions from the lower
hyperfine level, the model breaks down, demonstrating that factors
not included in the model play a significant role.

The layout of this paper is as follows.  In section 2 we derive
expressions needed to construct our model, present results for the
time-dependence of the population of relevant hyperfine states, and
discuss how spectra are produced. Section 3 describes the
experimental apparatus used to test the model's predictions in
rubidium and cesium, and section 4 contains experimental and
theoretical spectra. Section 5 contains a discussion of our results,
and in section 6 we draw our conclusions.

\section{Theory}
\subsection{Optical Bloch and population change equations}
The density matrix is used to calculate the atom-light interaction.
Consider a two level atom with ground state $a$, excited state $b$,
and transition angular frequency $\omega_{0}$, interacting with
laser light of angular frequency $\omega_{\rm L}$ and detuning
$\Delta = \omega_{\rm L}-\omega_{0}$. The Optical Bloch equations
are ~\cite{CCT}

\begin{eqnarray}
\dot{\rho}_{bb} & = & -i\frac{\Omega_{\rm R}}{2} (\tilde{\rho}_{
ab}-\tilde{\rho}_{ba})
-\Gamma \rho_{bb} \label{one}\nonumber\\
\dot{\tilde{\rho}}_{ba} & = & i\Delta \tilde{\rho}_{ba}+
i\frac{\Omega_{\rm R}}{2}(\rho_{bb}-\rho_{aa})
-\frac{\Gamma}{2} \tilde{\rho}_{ba} \label{two}\nonumber\\
\dot{\tilde{\rho}}_{ab} & = & -i\Delta \tilde{\rho}_{ab}-
i\frac{\Omega_{\rm R}}{2}(\rho_{bb}-\rho_{aa})
-\frac{\Gamma}{2} \tilde{\rho}_{ab} \label{three}\nonumber\\
\dot{\rho}_{aa} & = & i\frac{\Omega_{\rm R}}{2} (\tilde{\rho}_{
ab}-\tilde{\rho}_{ba}) +\Gamma \rho_{bb}~. \label{eq:four}
\label{eq:obe}
\end{eqnarray}

Here $\rho_{aa}$ and $\rho_{bb}$ are the ground and excited state
probabilities, respectively, and the off-diagonal terms
$\tilde{\rho}_{ab}$ and $\tilde{\rho}_{ba}$ are coherences. (The
tilde notation means that we are using ``slow'' variables - i.e. the
evolution of the coherence at the laser frequency has been factored
out separately).  The Rabi frequency is denoted by $\Omega_{\rm R}$
and is proportional to the product of the electric field of the
laser and the transition matrix element. The effect of spontaneous
emission is to reduce the excited state population at a rate
$\Gamma=1/\tau$, and the coherences at a rate $\Gamma/2$, where
$\tau$ is the excited state lifetime~\cite{CCT}. There are no
collisional terms as the mean time between collisions in the room
temperature vapor cells used in these experiments is far longer than
the mean time spent in the laser beams.  Similar equations are
derived for many-level atoms for every Zeeman sub-state.

It becomes much easier to solve these equations if the coherences
can be eliminated. To show that such an elimination is valid for
our system, we numerically solved the optical Bloch equations
containing the coherences and the rate equations for a few sample
systems. For a short time the solutions differ; the rate equations
are linear in time whereas the optical Bloch equations give a
quadratic time dependence for the evolution of populations.
However, the solutions become indistinguishable after a time of
$\sim 7\tau$~\cite{Coppendale}, where $\tau$=26.24(4) ns for
rubidium \cite{Volz} and 30.57(7) ns for cesium \cite{Rafac}. As
the typical time of flight of an atom through the laser beams
(discussed below) is roughly two orders of magnitude longer
($\simeq 5 \times 10^{-6}$ s) it is an excellent approximation to
solve the equations for the rate of change of populations. We
therefore assume that the coherences evolve sufficiently quickly
that their steady state value can be used.

With the coherences eliminated, we set the time derivatives to zero
in the second and third equations of (\ref{eq:obe}) and substitute
into the first equation, obtaining

\begin{align}
\begin{split}\label{eq:rate}
\dot{\rho}_{bb} = -\frac{\Omega_{\rm R}^2}{\Gamma}\frac{
(\rho_{bb}-\rho_{aa})}{1+4\Delta^2/ \Gamma^2} -\Gamma \rho_{bb}\\
 = -\frac{\Gamma}{2}\frac{I}{I_{\rm sat}}\frac{
(\rho_{bb}-\rho_{aa})}{1+4\Delta^2/ \Gamma^2} -\Gamma \rho_{bb}.
\end{split}
\end{align}

This leaves rate equations for populations only. Recalling that the
light intensity $I$ is proportional to the square of the Rabi
frequency, the saturation intensity $I_{\rm sat}$ is defined as
$I/I_{\rm sat}=2\Omega_{\rm R}^2/\Gamma^2$. The three terms on the
right-hand side of (\ref{eq:rate}) have the usual physical
interpretation: the first represents stimulated emission out of the
excited state, which is proportional to the light intensity; the
second absorption into the excited state, also proportional to the
light intensity; and the third spontaneous emission out of the
excited state, which is independent of the light intensity.  The
intensity-dependent terms have a Lorentzian line-shape, with full
width at half maximum of $\Gamma$.  Note that the usual rate
equations of laser physics for populations with Einstein A and B
coefficients assume broad-band radiation, and use the intensity per
bandwidth (evaluated at line center); in contrast, equation
(\ref{eq:rate}) is valid for narrow-band radiation and is a function
of the detuning.

It is straightforward to generalize these equations for multilevel
systems. As there are three possible polarizations, and the $\Delta
F=0, \pm1$ selection rule applies, each excited state can decay into
up to nine different ground states.  The pump beam drives
$\sigma^{+}$ transitions, and the Rabi frequencies and saturation
intensities are calculated from the known line-strength
coefficients.

The line strength for a transition between two states $a$ and $b$
is proportional to the square of the dipole matrix element, given
\cite{Edmonds} by

\begin{eqnarray}
\vert d_{ab} \vert^2 & = & \vert e\langle n_b L_b \vert \vert r
\vert \vert n_a L_a \rangle\vert^2 \\
& \times & (2J_b+1)(2J_a+1)(2F_b+1)(2F_a+1)~~~~~~~ \nonumber \\
& \times & \left[\left\{\begin{array}{ccc}
L_b& J_b& S\\
J_a& L_a& 1\\
\end{array}
\right\} \left\{\begin{array}{ccc}
J_b& F_b& \mathcal{I}\\
F_a& J_a& 1\\
\end{array}
\right\} \left( \begin{array}{ccc}
F_a& 1& F_b\\
m_{F_{a}}& q& -m_{F_{b}}\\
\end{array}
\right)\right]^2~,\nonumber
\end{eqnarray}

where $L$ and $S$ are respectively the orbital and spin angular
momenta of the electron, $J$ is their sum, $F=\mathcal{I}+J$ where
$\mathcal{I}$ is the nuclear spin, and $m_F$ is the projection of
$F$ onto the $z$ axis. The expressions in curly brackets are
standard $6J$ symbols encountered when re-coupling angular momenta,
and the term with large round brackets is a $3J$ symbol. These can
be calculated using standard symbolic mathematical packages
\cite{Mathematica}. The value of $q$ denotes the change in
$z$-component of the angular momentum in the transition between the
two levels, with $q=0$ for $\pi$ transitions, and $q=\pm 1$ for
$\sigma^{\pm}$ transitions.

As an example, consider $^{87}$Rb ($\mathcal{I}=3/2$), which has two
hyperfine states in the ground term, $F=2,1$. If the pump laser is
tuned in the vicinity of the resonances $S_{1/2}(F=2)\rightarrow
P_{3/2}(F'=3,2,1)$ with the detuning, $\Delta$, defined relative to
the closed transition $S_{1/2}(F=2)\rightarrow P_{3/2}(F'=3)$, the
strongest transition will be
$^{2}S_{1/2}(F=2,m_F=2)\rightarrow$$^{2}P_{3/2}(F'=3,m_{F'}=3)$. For
this transition, the saturation intensity $I_{\rm sat}$ has the
value 1.6~mW/cm$^{2}$.  The reduced matrix element for the
transition $\langle 5S \vert \vert r \vert \vert 5P \rangle = 5.14
a_0$,  where $a_0$ is the Bohr radius, is calculated from the
excited state lifetime. The line-strength ratios, which we define as
\begin{equation}
R_{F,m_F\rightarrow F',m_{F'}}=\left( \frac{d_{F,m_F\rightarrow
F',m_{F'}}}{d_{2,2\rightarrow3,3}}\right)^2 , \label{eq:ratio}
\end{equation}
are always less than one as they are normalized relative to the
strongest transition.

To simplify our notation we write the population of a state as
$P_{F,m_F}$, which is equal to the diagonal density-matrix element
$\rho_{F,m_F,F,m_F}$. The five $F=2, m_F$ ground states obey the
rate equations
\begin{displaymath}
\frac{dP_{F,m_F}}{dt}  =  -\sum_{F' = F-1}^{F' = F+1}
R_{F,m_F\rightarrow F',m_F+1}\frac{\Gamma}{2}\frac{I}{I_{\rm
sat}}\frac{ (P_{F,m_F}-P_{F',m_{F+1}})}{1+4\left(\Delta
'/\Gamma\right)^2}
\end{displaymath}

\begin{equation}
+\sum_{m_{F'}=m_F -1}^{m_{F'}=m_F +1} \sum_{F'=F-1}^{F'=F+1}
R_{F,m_F\rightarrow F',m_{F'}}\Gamma P_{F',m_{F'}},
\label{eq:rateupper}
\end{equation}
where $\Delta'=\Delta$ if $F'=F+1$; $\Delta'=\Delta+\Delta_{32}$ if
$F'=F$; and $\Delta'=\Delta+\Delta_{31}$ if $F'=F-1$. Here $\hbar
\Delta_{32}$ and $\hbar \Delta_{31}$ are the excited state hyperfine
intervals. The first row of Eq. (\ref{eq:rateupper}) contains the
stimulated absorption and emission terms, the second row the
spontaneous emission terms.

The three $F=1, m_F$ ground states obey the rate equations

\begin{equation}
\frac{dP_{F,m_F}}{dt}  =\sum_{m_{F'}=m_F -1}^{m_{F'}=m_F +1}
\sum_{F'=F-1}^{F'=F+1} R_{F,m_F\rightarrow F',m_{F'}}\Gamma
P_{F',m_{F'}}. \label{eq:ratelower}
\end{equation}
Note that there are no stimulated terms as the large ground state
splitting means that these transitions very far off resonance; these
states can only increase their population by spontaneous emission
from the excited states.

Finally, the excited state population rate equations are
\begin{displaymath}
\frac{dP_{F',m_{F'}}}{dt}  =  \sum_{F = 2} R_{F,m_{F'}-1\rightarrow
F',m_{F'}}\frac{\Gamma}{2}\frac{I}{I_{\rm sat}}\frac{
(P_{F,m_{F'}-1}-P_{F',m_{F'}})}{1+4\left(\Delta '/\Gamma\right)^2}
\end{displaymath}

\begin{equation}
-\sum_{m_{F}=m_{F'} -1}^{m_{F}=m_{F'} +1} \sum_{F=F'-1}^{F=F'+1}
R_{F,m_F\rightarrow F',m_{F'}}\Gamma P_{F',m_{F'}}.
\label{eq:rateexcited}
\end{equation}

A useful check before solving these equations is that the sum of all
of the right-hand sides of these equations must be zero, because the
total population in all states is constant. Note that there are 36,
24, or 48 Zeeman levels which have to be considered for the
$^{85}$Rb, $^{87}$Rb, and $^{133}$Cs systems, respectively; hence
the simplification of the equations by eliminating the coherences is
substantial.

\subsection{Calculation of anisotropy}

The birefringence of the medium can be probed by a linearly
polarized probe beam, which can be decomposed into two beams of
equal amplitude and opposite circular polarization.  For an
isotropic medium, both circular components experience the same
refractive index and absorption.  There is a difference in
absorption coefficients, $\Delta \alpha$, for an optically pumped
medium: $\Delta \alpha = \alpha_{+}-\alpha_{-}$, with $\alpha_{\pm}$
being the absorption coefficients of the circular components driving
$\sigma^{\pm}$ transitions. Correspondingly, the incident linearly
polarized beam exits the medium with a rotated elliptical
polarization. Both the ellipticity and the angle of rotation are
proportional to $\Delta \alpha$ \cite{Pearman}.  In our experimental
set-up (Section 3) we are sensitive to the rotation of polarization,
providing the absorption coefficients are not too large. The signal
has a characteristic dispersion spectrum~(equation (9) in
\cite{Pearman}). Therefore for each resonance we calculate one
parameter, the line-center difference in absorption.

The experiments were performed with alkali metal atoms ($^{85}$Rb,
$^{87}$Rb, $^{133}$Cs) on the D2 transition.  For an atom in the
$^{2}S_{1/2}$ ground state with nuclear spin  $\mathcal{I}$ there
are two values for the total angular momentum $F$, namely
$F=\mathcal{I}\pm1/2$. There are four values for the excited state
angular momentum $F'$, namely $F'=\mathcal{I}\pm3/2,
\mathcal{I}\pm1/2$. The hyperfine splitting of the ground states
(3.0 GHz for $^{85}$Rb, 6.8 GHz for $^{87}$Rb, 9.2 GHz for
$^{133}$Cs) exceeds the room temperature Doppler width ($\sim$~0.5
GHz), whereas the excited state hyperfine splitting is less than the
Doppler width.  Therefore, the absorption spectrum consists of two
isolated Doppler broadened absorption lines per isotope. Sub-Doppler
absorption features are obtained by using counter-propagating pump
and probe beams. There are three excited state resonances coupled to
each ground state via electric dipole transitions, namely
$S_{1/2}(F)\rightarrow P_{3/2}(F'=F+1,F,F-1)$. In addition, one
observes three cross-over resonances at frequencies halfway between
each pair of conventional resonances~\cite{Macadam92}.

The calculation assumes that there is no excited state population
initially, and that the ground state population is spread uniformly
amongst the $2(2\mathcal{I}+1)$ different $|F,m_F\rangle$ levels.
The pump beam has intensity $I$ and drives $\sigma^{+}$ transitions.
This optical pumping process drives the population towards the
largest possible value of $m_{F}$, and the medium becomes optically
anisotropic. Equations similar to (\ref{eq:rate}) are written down
for each $m_F$ level, and the set of coupled equations are solved
numerically.

\begin{figure}
\centering
\includegraphics[width=8.5cm]{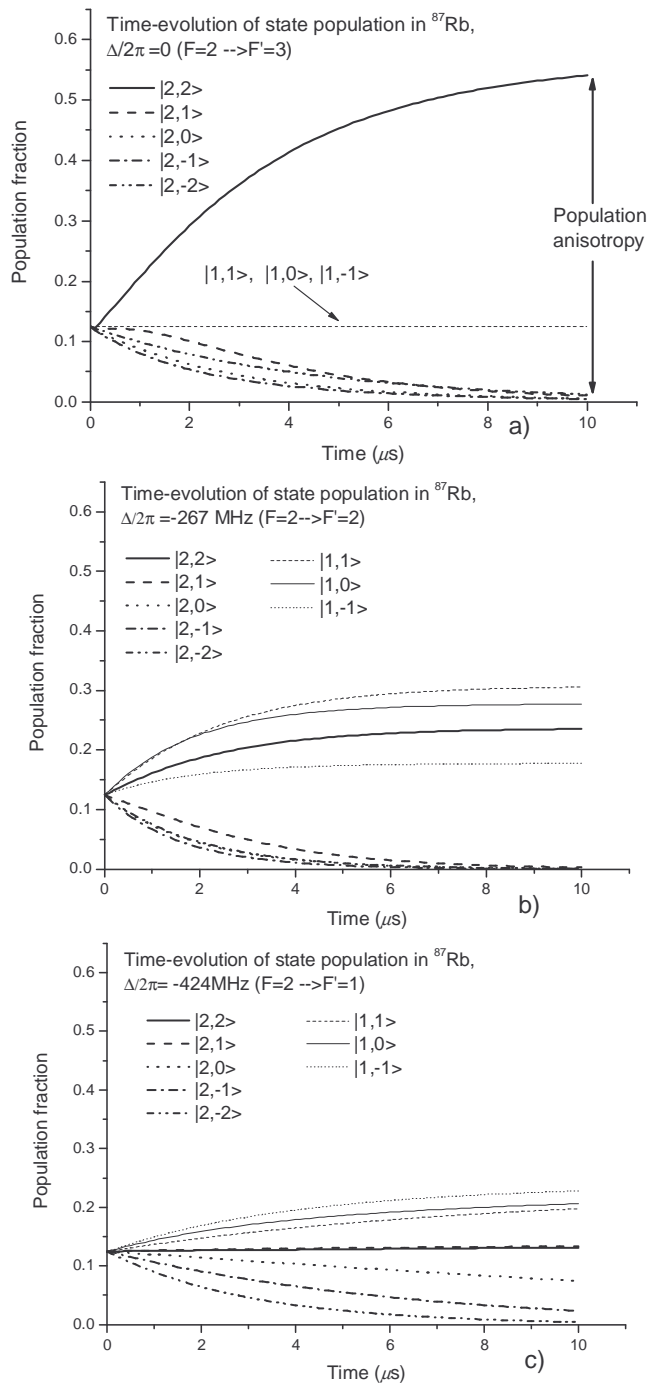}
\caption{Time-evolution of population. The population of the eight
ground state sub-levels of $^{87}$Rb are plotted as a function of
time after the pump beam is turned on. The pump drives $\sigma^+$
transitions and has an intensity of 0.1$I_{\rm{sat}}$. Vertical axes
are identical to emphasize differences in population behavior. a)
Laser tuned to $F=2 \rightarrow F'=3$ transition. Within a few
$\mu$s the initial isotropic population distribution becomes highly
anisotropic, as atoms are optically pumped into the $|2,2\rangle$
state.  As this transition is closed, no change occurs in the
population of the $F=1$ manifold. b) When the laser is tuned to the
$F=2 \rightarrow F'=2$ transition the atoms are again optically
pumped towards the $|2,2\rangle$ state. However, as this is an open
transition, a significant fraction of population accumulates in the
$F=1$ manifold. These states do not contribute to the medium's
optical anisotropy. c) For the laser turned to the $F=2 \rightarrow
F'=1$ transition even fewer atoms are pumped into the $|2,2\rangle$
state. } \label{population}
\end{figure}

From these solutions, we obtain graphs of the population of each
hyperfine state as a function of time. Figure \ref{population} shows
the time-evolution of these populations for a sample of $^{87}$Rb
atoms when $\Delta/2\pi$=0, -267 MHz, and -424 MHz, corresponding to
a laser on resonance with the $F=2 \rightarrow F'=3,2,1$ transitions
respectively. When the laser is resonant with the closed $F=2
\rightarrow F'=3$ transition, atoms are rapidly pumped into the
extreme $|2,2\rangle$ state, with over 50\% of atoms in this state
after $<$10 $\mu$s. The population of the other states is either
constant or decreasing after $\sim$ 1 $\mu$s.

These figures clearly show the different optical pumping dynamics as
a function of the laser detuning. A detuning $\Delta$=0 (Figure
\ref{population}a) is optimum for achieving the most optically
anisotropic medium. For $\Delta/2\pi$=-267 MHz (Figure
\ref{population}b) the medium becomes optically anisotropic, but to
a lesser extent. This is due to the open nature of the transition,
which allows atoms to accumulate in the $F=1$ manifold, where they
are too far from resonance to influence the medium's anisotropy.
This effect is even more pronounced for $\Delta/2\pi$=-424 MHz
(Figure \ref{population}c), and the medium's optical anisotropy at
this detuning is correspondingly reduced in comparison with the
other detunings.

\begin{figure}[htb]
\centering
\includegraphics[width=8.5cm]{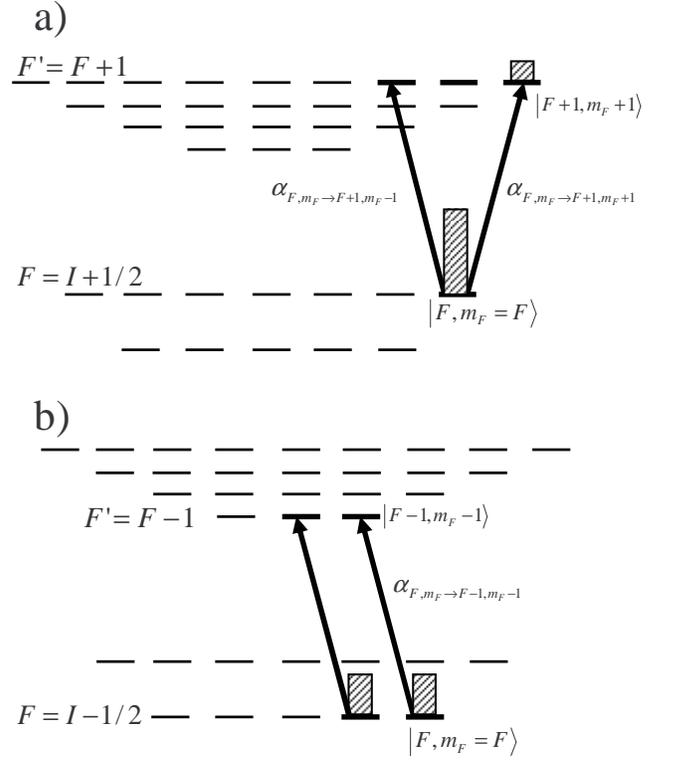}
\caption{Sign of anisotropy for closed transitions. a) For the
closed $F=\mathcal{I}+1/2 \rightarrow F'=F+1$ transition the
population is optically pumped into the $|F,m_F=F\rangle$ state,
with a small fraction in the excited state. The line strength
$\alpha_{F,m_F \rightarrow F+1,m_F+1}$ is significantly larger than
$\alpha_{F,m_F \rightarrow F+1,m_F-1}$. b) For the closed
$F=\mathcal{I}-1/2 \rightarrow F'=F-1$ transition the population is
optically pumped into the $|F,m_F=F\rangle$ and $|F,m_F=F-1\rangle$
states. There are no allowed $\sigma^+$ transitions for these
states, whereas the line strength for the $\sigma^-$ transitions are
finite.  Consequently, the anisotropy of the medium has opposite
sign relative to a). All levels are drawn as for an atom with
$\mathcal{I}=5/2$; the structure of the other alkali atoms is
similar.} \label{distribution}
\end{figure}

These population graphs allow us to predict that the polarization
spectroscopy signal should be dominated by the closed transitions,
$S_{1/2}(F=\mathcal{I}+1/2)\rightarrow P_{3/2}(F'=F+1)$ and
$S_{1/2}(F=\mathcal{I}-1/2)\rightarrow P_{3/2}(F'=F-1)$. For such
transitions, selection rules forbid atoms in the excited state
from falling into the other ground state; consequently, all of the
ground state population ends up pumped into the extreme magnetic
sublevels with $m_{F}=F$ (for $\mathcal{I}+1/2$, as discussed in
the example above) or $m_{F}=F$ and $m_{F}=F-1$ (for
$\mathcal{I}-1/2$). The absorption coefficients $\alpha_{\pm}$
differ most for this extreme state; therefore, the anisotropy of
the medium is maximized. Note that the sign of the anisotropy
generated by the $\mathcal{I}+1/2$ transitions is expected to be
opposite that of the lower hyperfine transitions, due to the lack
of allowed $\sigma^+$ transitions from the ground
$m_F=F=\mathcal{I}-1/2$ states. Figure \ref{distribution}
illustrates the optical pumping process.

After solving the coupled population rate equations it is possible
to calculate the time-dependent anisotropy of the medium, $\mathcal
A(t)$.  This is defined as a sum over all of the ground $m_{F}$
states of the difference in the absorption coefficients for the
components of the probe beam driving $\sigma^{\pm}$ transitions,
taking into account the ground and excited state populations:

\begin{displaymath}
\mathcal{A}(t) = \sum_{m_{F}=-F}^{m_{F}=+F}
\alpha_{(F,m_{F}\rightarrow F',m_{F+1})}\left(P_{F,m_{F}}-P'_{F',
m_{F}+1}\right)
\end{displaymath}
\begin{equation}
-\alpha_{(F,m_{F}\rightarrow F',m_{F-1})}\left(P_{F,m_{F}}-P'_{F,
m_{F}-1}\right).
\end{equation}

For each isotope and each ground state $F$ the anisotropy is
calculated for the three frequencies corresponding to the resonances
$F\rightarrow (F'=F+1, F, F-1)$, i.e. $\Delta$=0, $-\Delta_{32}$,
and $-\Delta_{31}$. It is assumed that since the experiment uses
counter propagating pump and probe beams only atoms with zero
velocity along the axis of the beams contribute significantly. The
details of the transverse motion of the atoms is outlined in the
next sub-section.

\subsection{Transverse motion of atoms}

The experiment measures the average anisotropy of the medium. To
simulate this, we average the time-dependent anisotropy,
$\mathcal{A}(t)$, with a weighting function, $\mathcal{H}(t)$, which
gives the distribution of times of flight transverse to the beam.
For a circular beam of radius $a$ the probability distribution,
$\mathcal{F}(\ell)$, of having a path length $\ell$ in a uniform gas
is
\begin{equation}
\mathcal{F}(\ell) = \frac{\ell}{2a\sqrt{4a^2-\ell^2}}.
\end{equation}
The probability distribution function, $\mathcal{G}(t,\ell)$,  for
having a transit time $t$ for a given $\ell$ is obtained from the
Maxwell velocity distribution for a sample of atoms of mass $m$ at
temperature $T$ and is
\begin{equation}
\mathcal{G}(t,\ell) = \frac{m\ell^2}{k_{\rm B}Tt^3}\exp\left(
-\frac{m\ell^2}{2k_{\rm B}Tt^2}\right).
\end{equation}
The distribution function $\mathcal{H}(t)$ is obtained thus:
\begin{equation}
\mathcal{H}(t) = \int_{\ell=0}^{2a}
\mathcal{G}(t,\ell)\mathcal{F}(\ell)d\ell,
\end{equation}
which can either be evaluated numerically or in closed form in terms
of the complex error function. The average anisotropy
$\overline{\mathcal{A}}$ is calculated by averaging the
time-dependent anisotropy with the time of flight weighting
function, i.e.:
\begin{equation}
\overline{\mathcal{A}}=\int
\mathcal{A}(t)\mathcal{H}(t)dt.\label{eq:anisotropy}
\end{equation}
This integral is calculated numerically at the three resonant
frequencies of each D2 transition.
\subsection{Generating theoretical spectra}

To generate the predicted spectra, the three values of
$\overline{\mathcal{A}}$ from (\ref{eq:anisotropy}) are multiplied
by dispersion functions of the form $x/(1+x^2)$, where
$x=2\Delta/\Gamma$. As in the work of Yoshikawa {\it et
al.}~\cite{Yoshikawa} the strength of cross-over features is assumed
to be the average of the two associated resonances; hence cross-over
features are calculated by multiplying this average by a dispersion
function located halfway between each resonance.

To account for spectral broadening produced by saturation effects
(power broadening), we substituted a broadened linewidth $\gamma =
k\Gamma$ MHz for the natural linewidth $\Gamma$ in the dispersion
function. We found that for $1\leq k\leq 2$, the experimental data
do not provide a tight constraint on the value of $\gamma$.
Doppler broadening is incorporated by convolving the resulting
spectra with a Gaussian of FWHM $\leq 1 \times 2\pi$ MHz,
consistent with residual Doppler broadening due to the finite
crossing angle between the probe and pump beams. Magnitudes for
both types of broadening were set by the experimental conditions,
as were the values for temperature (293K for rubidium, 273K for
cesium), beam radii, and $I/I_{\rm sat}$.

\section{Experimental setup}
\begin{figure}
\centering
\includegraphics[width=8.5cm]{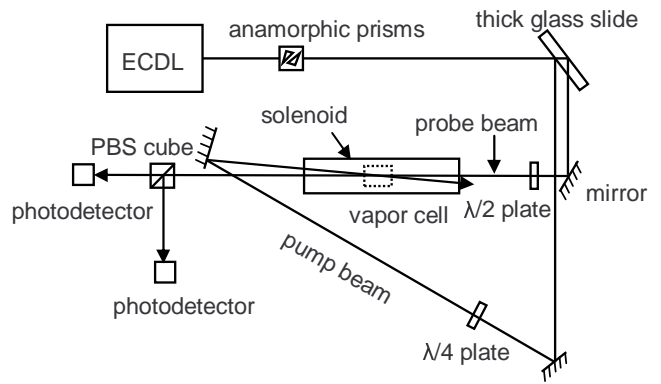}
\caption{Experimental layout. A thick glass slide picks off a
fraction of the extended cavity diode laser (ECDL) light and splits
it into two parallel beams.  One beam acts as the probe, and the
other as a (nearly) counterpropagating pump. The beams overlap
inside a 70 mm (50 mm) Rb (Cs) cell, which rests inside a long
solenoid and below a coil that cancels the ambient laboratory field.
In the cesium experiment two coils in the Helmholtz configuration
replace the solenoid.  A half-wave plate rotates the plane of
polarization of the probe beam with respect to the polarizing beam
splitter (PBS) cube axis; a quarter-wave plate makes the pump
circularly polarized.} \label{layout}
\end{figure}

The layout of our experiment is similar to that described in
\cite{Pearman} and is shown in Figure \ref{layout}. The probe
beam's plane of polarization is set at $\pi$/4 with respect to a
polarizing beam splitter (PBS), which acts as an analyzer. The
signal we record is the difference between the signals in each arm
of the PBS; in the absence of a pump beam, the two arms will have
equal intensities, and the difference will be zero. This technique
produces polarization signals an order of magnitude larger than
the conventional (single detector) method of polarization
spectroscopy - an important advantage if the signal is being used
for laser locking \cite{Pearman}.

The experimental layouts for rubidium and cesium spectroscopy are
very similar, differing only in the length of the alkali vapor cell,
equipment used to control the magnetic field along the axis of the
cell, and waveplate type. Table 1 contains information on the spot
radii ($1/e^2$ intensities) of the beams after the light has passed
through a pair of anamorphic prisms.

In both experiments the pump and probe beams are derived from the
same extended cavity diode laser (ECDL).  The diode for the rubidium
(cesium) laser was a Sanyo DL-7140-201 (SDL-5401-G1). The crossing
angle between probe and counterpropagating pump within the vapor
cell is $<$3.0$\pm$0.2~mrad. Neutral density filters are used to
vary pump and probe powers independently.  A half-wave plate rotates
the polarization of the probe relative to the axis of the PBS; a
quarter-wave plate converts light in the pump beam to circular
polarization. The two output beams from the PBS are focused onto
photodiodes, which are connected to simple current-to-voltage
circuits designed to output a voltage linearly proportional to the
incident intensity. These voltages are then subtracted
electronically to yield the polarization spectra.
\begin{table}
\caption{\label{laserspecs} Specifications of lasers used in this
work.}
\begin{center}
\begin{tabular}{c|c|c|c}
\hline \hline Atom & $\lambda$ & Horizontal $r$ ($1/e^2$) & Vertical
$r$
($1/e^2$)\\
\hline
Rb & 780 nm & 0.65$\pm$0.01 mm& 0.59$\pm$0.01 mm\\
\hline
Cs & 852 nm & 0.69$\pm$0.01 mm& 0.70$\pm$0.01 mm\\
\hline \hline
\end{tabular}
\end{center}
\end{table}

Since light of a given polarization may drive $\sigma^+$,
$\sigma^-$ or $\pi$ transitions depending on the external magnetic
field, it is necessary to establish a ``preferred" magnetic field
direction along the vapor cell axis. In the rubidium experiment
this is done by placing a room-temperature cell containing
$^{85}$Rb and $^{87}$Rb inside a 300-turn solenoid of length 280
mm and diameter 26 mm. Numerical simulations showed that the
magnetic field inside the solenoid is uniform to within 0.2\% over
the length of the 70 mm cell.

In the cesium experiment the cell is partially submerged in an ice
bath, and two 300 $\times$ 300 mm square coils in the Helmholtz
configuration generate an axial magnetic field uniform to 1\%. The
ice bath is needed because our model assumes an optically thin
medium, but at 23 $^\circ$C absorption exceeds 90\% for a 50 mm Cs
vapor cell (compared to a maximum of 30\% for Rb in a 70 mm cell).
In the ice bath this is reduced to 60\% (50\%) for transitions from
the upper (lower) hyperfine level of the ground state.  In both
experiments we also cancel the (primarily vertical) ambient
laboratory magnetic field with a 245 mm diameter coil mounted above
the cell. Any inhomogeneity due to the finite diameter of this coil
makes a negligible contribution to the total field when added in
quadrature with the axial field.  The magnitude of the axial field
is set to a value just below the point where Zeeman splitting of the
hyperfine levels begins to distort the polarization spectra,
typically $\simeq 150$ $\mu$T.

\begin{figure}
\centering
\includegraphics[width=8.5cm]{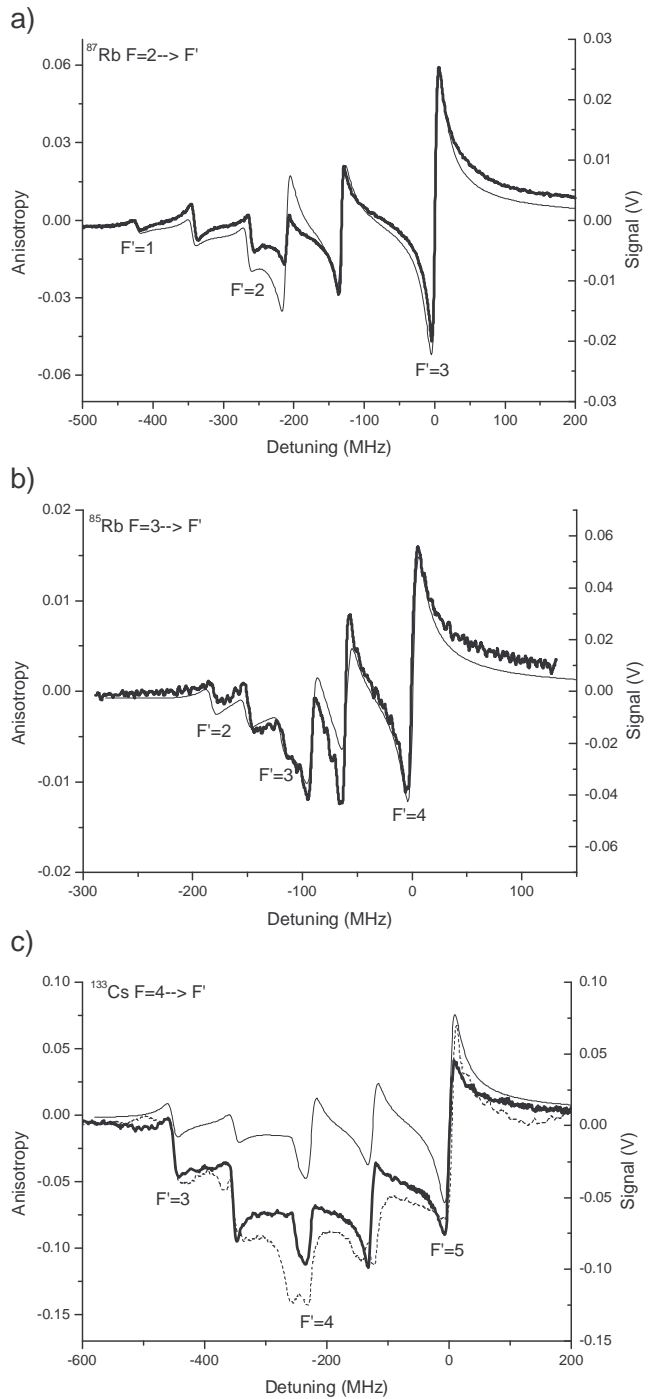}
\caption{Experimental (thick line) and theoretical (fine line)
polarization spectra of D2 line transitions in a) $^{87}$Rb, b)
$^{85}$Rb, and c) $^{133}$Cs. Spectra were obtained by tuning the
laser frequency to drive transitions from the upper hyperfine level
of the atom's ground state. All three spectra are dominated by the
strong dispersion features associated with the closed transition.
For the Cs upper hyperfine spectra, solid (dashed) lines represent
spectra taken with the vapor cell at 0$^\circ$ (23$^\circ$)
C.}\label{experimentandtheory}
\end{figure}

\section{Results}
Figure \ref{experimentandtheory} shows the data obtained with the
layout described in the last section (thick line) and the
theoretical spectra (thin line). All experimental spectra were taken
with pump and probe beam intensities $\leq$ 0.1 $I_{\rm sat}$ to
reduce saturation effects. The spectra shown are of transitions from
the upper hyperfine level of the ground state, i.e. the $5
\:^2S_{1/2}\: (F = 2) \rightarrow 5\: ^2P_{3/2} \:(F' = 1,2,3)$
transitions in $^{87}$Rb, $5\: ^2S_{1/2} \:(F = 3) \rightarrow 5
\:^2P_{3/2} \:(F' = 2,3,4)$ transitions in $^{85}$Rb, and $6
\:^2S_{1/2} \:(F = 4) \rightarrow 6 \:^2P_{3/2} \:(F' = 3,4,5)$
transitions in $^{133}$Cs. The closed transitions in this group are
the well-known ``cooling" transition used in laser-cooling
experiments. Zero detuning is relative to the highest-frequency
transition, and all detunings are given in units of $\Delta / 2\pi$.
The magnitude of each feature is given in volts, and will depend on
the gain resistance in the photodiode circuit (1 M$\Omega$ in our
experiment). For Cs, two sets of data are shown; one taken with the
vapor cell at room temperature (dashed line), the other with the
cell in an ice bath as described in the previous section (solid
line).

\begin{figure}[htb]
\centering
\includegraphics[width=8.5cm]{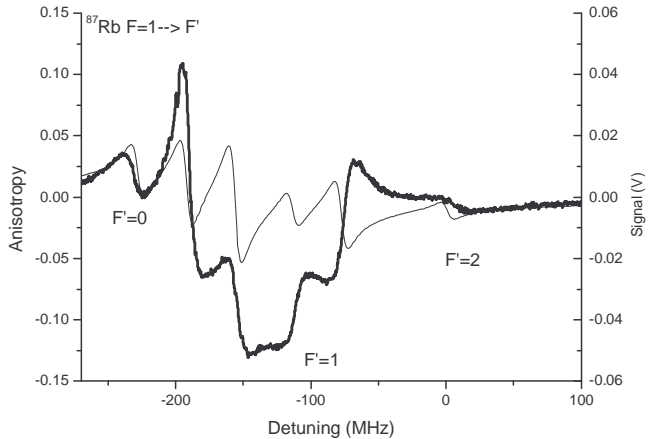}
\caption{Experimental (thick line) and theoretical (fine line)
polarization spectra of D2 line transitions from the lower hyperfine
level of the ground state in $^{87}$Rb.}\label{Rb87Repump}
\end{figure}

In these spectra, polarization signals of closed transitions
dominate, with magnitudes up to three times that of the next
largest feature. This is true for all three species, and is in
marked contrast with conventional saturated absorption/hyperfine
pumping spectra \cite{Smith}. As predicted, the $F \rightarrow
F'=F$ transitions produce a moderate amount of anisotropy, and the
lowest frequency open transitions give the smallest signal.  A
graph showing the distribution of times of flight for rubidium at
room temperature through a beam with the experimental waist shows
a most probable time of flight of 5 $\mu$s, and a significant
fraction of the atoms with times of flight longer than 10 $\mu$s.
Therefore we do indeed see that the timescale needed to build up
anisotropy is significantly longer than the spontaneous decay
time, and comparable to transit times. The accumulation of
anisotropy can be accelerated by using a more intense pump beam;
however, this will not greatly increase the signal as most atoms
are already pumped into the extreme state within the transit time.
This is also what is observed experimentally.

Spectra taken of transitions from the lower hyperfine level of the
ground state show a more complicated pattern. The (low-frequency)
closed transitions in Cs and $^{85}$Rb show a characteristic strong
dispersion feature, but in $^{87}$Rb the largest feature arises from
one of the cross-over peaks (Figure \ref{Rb87Repump}). In $^{85}$Rb
closely-spaced transitions generate a polarization spectrum in which
individual peaks merge, making exact matching of features and
transitions difficult.

\section{Discussion}
By comparing theoretical and experimental traces, we see immediately
that our model reproduces experimental features of upper hyperfine
transition spectra with a high degree of accuracy. Fine spectral
details like the ``horns" resulting from the closely-spaced $F = 2
\rightarrow F' = 2$ and $X_{3,1}$ cross-over peak in $^{87}$Rb arise
automatically from the calculated anisotropies. The magnitudes of
theoretical spectral peaks relative to the large closed transition
peaks also agree well with experimental data, especially in
$^{85}$Rb. The effect of reducing vapor pressure by cooling the Cs
cell is readily apparent. Although the central peaks are still
offset from theoretical spectra compared to their counterparts in
the Rb spectra, the magnitude of the offset is significantly less at
0 $^\circ$C than at room temperature, and the shapes of the spectra
broadly agree.

For lower hyperfine spectra (e.g. the $^{87}$Rb spectra in Figure
\ref{Rb87Repump}), we observe that the sign of the anisotropy for
the closed-transition peak is reversed compared to corresponding
peaks in upper hyperfine spectra, as predicted in our model.
However, most other spectral features differ markedly from
predictions. Crucially, though, the model spectra show similar
discrepancies for each of the three species, indicating that the
lack of agreement must be due to physical processes not included in
our model. As a test, we ran our simulation again, this time
allowing the position and magnitude of the three Lorentzians
(representing three transitions) to float unconstrained by any input
parameters.  We found that no values of detuning or anisotropy could
account for the observed spectra.

To gain insight into the anisotropies induced by lower hyperfine
transitions (particularly ``open" transitions, i.e. those with a
decay channel to the other ground state level), we performed
experiments designed to measure the anisotropy directly. As in
Reference \cite{Pearman}, a quarter-wave plate was inserted before
the PBS, oriented such that it converted the circular polarization
component of the probe which drives $\sigma^+$ ($\sigma^-$)
transitions into vertically (horizontally) polarized light. The
output from the arms of the PBS then directly measures the
absorption experienced by the components driving $\sigma^+$ and
$\sigma^-$ transitions.

For upper hyperfine transitions, the largest anisotropies (i.e.
largest difference in signal between the two arms) resulted from the
closed transitions.  This agrees with theoretical predictions and is
reflected in our data. For lower hyperfine transitions, large and
strongly negative anisotropies occurred at the cross-over
frequencies - a fact reflected in our data, but not in the model.
This suggests that for such transitions, it is no longer valid to
assume that the strength of crossover features is the average of the
two associated resonances.

The present model assumes a uniformly distributed circular laser
beam.  The experimental beam has a Gaussian profile, which will
lead to slight variations in the absolute height of the
spectroscopic peaks. It will not, however, account for the
difference between upper and lower hyperfine transitions. It is
also possible that if the optical pumping is not complete for the
lower hyperfine states, ignoring the coherences could have a large
effect.  The most significant approximation that is likely to
break down for lower hyperfine transitions is the assumption that
only atoms with no axial velocity will contribute to the
anisotropy. As was shown in \cite{Smith}, a full description of
the spectrum, especially open transitions, must take into account
a large velocity class of atoms moving along the beam, not just
the ``stationary" ones.  The contribution of these nonzero
velocity classes will be more significant for $F=\mathcal{I}-1/2$
because the line strength factors for the closed transition are
weakest; for $F=\mathcal{I}+1/2$, line strength factors are
strongest for the closed transition, and the ``stationary atom"
approximation is reasonable.

Expanding the model to include a large velocity class would be
computationally intense, and is beyond the scope of this work.
Walewski et al. have performed numerical simulations which included
nonzero axial velocities, and successfully used their model to
explain features of polarization spectra in flames \cite{Walewski}.
However, the organic molecules they studied have very different
structures and properties from those of the alkali atoms examined
here, and adapting their model to account for alkali spectra would
be correspondingly nontrivial.

In comparing theoretical predictions with experimental signals, we
have chiefly focused on the shape and relative magnitudes of
spectral features. The absolute magnitude of the experimental
polarization spectroscopy signal depends not only on the gain
resistor in the photodiode circuit, as discussed in the previous
section, but also on the intensity of the probe beam in the absence
of a vapor cell (see Eq. 7 in reference \cite{Pearman}). This in
turn depends on the optical thickness of the vapor, which is both
temperature-dependent (as noted earlier in our discussion of cesium)
and different for each species. Because of these factors, the
species with the largest calculated anisotropy will not necessarily
produce the largest experimental signal.  Our decision to display
theoretical and experimental spectra on separate scales reflects
these considerations.

Finally, we note that although the peaks associated with the closed
transitions in the upper hyperfine spectra are ideal for laser
locking, with steep slopes centered on line center, this is not true
for the lower hyperfine spectra, where the position of each zero
crossing is not trivially related to the position of the resonances.

\section{Conclusions}

We have presented a model of polarization spectroscopy based on
numerical integration of population rate equations, and shown how
this model can be used to simulate the optical pumping process that
leads to an anisotropic population distribution, and thus an
optically anisotropic medium. Theoretical polarization spectra
generated by this model account very well for $F=\mathcal{I}+1/2
\rightarrow F'$ transitions in $^{87}$Rb, $^{85}$Rb, and $^{133}$Cs,
but not as well for $F=\mathcal{I}-1/2 \rightarrow F'$ transitions.
For $F=\mathcal{I}+1/2 \rightarrow F'$, the closed $F'=F+1$
transition dominates the spectra, with a steep slope through line
center which makes an ideal frequency reference for laser locking.
Although we have not studied them experimentally in this work, we
would expect similar results for the D2 spectra of other alkali
species, e.g. $^{6}$Li and $^{23}$Na.

\section*{Acknowledgements}

We thank V. Jacobs and R. C. Shiell for valuable discussions. We
acknowledge support from the UK EPSRC, the Universities UK Overseas
Research Scheme (MLH) and the Royal Society (SLC).

\end{document}